# Tailoring magnetization reversal of a single-domain bar nanomagnet via its end geometry

Jianhua Li[1,2], Sining Dong[1,*], Wen-Cheng Yue[1], Zixiong Yuan[1], Zhi-Li Xiao[3,4], Yang-Yang Lyu[1], Ting-Ting Wang[1], Chong Li[1], Chenguang Wang[1], Wen-Bing Xu[1], Ying Dong[5], Huabing Wang[1,6], Peiheng Wu[1], Wai-Kwong Kwok[3] and Yong-Lei Wang[1,*]

[1] *Research Institute of Superconductor Electronics, School of Electronic Science and Engineering, Nanjing University, Nanjing, 210023, China*

[2] *School of Physics and Electronic Electrical Engineering, Huaiyin Normal University, Huaian, 223300, China*

[3] *Materials Science Division, Argonne National Laboratory, Argonne, IL 60439, USA*

[4] *Department of Physics, Northern Illinois University, DeKalb, IL 60115, USA*

[5] *Research Center for Quantum Sensing, Zhejiang Lab, Hangzhou, Zhejiang, 311121, China*

[6] *Purple Mountain Laboratories, Nanjing, 211111, China*

\* Correspondence to: sndong@nju.edu.cn; yongleiwang@nju.edu.cn

**Abstract**

Nanoscale single-domain bar magnets are building blocks for a variety of fundamental and applied mesoscopic magnetic systems, such as artificial spin ices, magnetic shape-morphing microbots as well as magnetic majority logic gates. The magnetization reversal switching field of the bar nanomagnets is a crucial parameter that determines the physical properties and functionalities of their constituted artificial systems. Previous methods on tuning the magnetization reversal switching field of a bar nanomagnet usually rely on modifying its aspect ratio, such as its length, width and/or thickness. Here, we show that the switching field of a bar nanomagnet saturates when extending its length beyond a certain value, preventing further tailoring of the magnetization reversal via aspect ratios. We showcase highly tunable switching field of a bar nanomagent by tailoring its end geometry without altering its size. This provides an easy method to control the magnetization reversal of a single-domain bar nanomagnet. It would enable new research and/or applications, such as designing artificial spin ices with additional tuning parameters, engineering magnetic microbots with more flexibility as well as developing magnetic quantum-dot cellular automata systems for low power computing.

## I. INTRODUCTION

Due to shape anisotropy, an elongated magnet with submicrometer dimensions (bar nanomagnet) is in a single-domain state with a bistable remnant magnetization pointing along its long axis[1]. These single-domain bar nanomagnets have been extensively used in a wide range of fundamental and applied mesoscopic magnetic systems. For example, coupled bar nanomagnets were used to design majority logic gates for low dissipation digital computation in magnetic quantum-dot cellular automata systems[1]. In artificial spin ices, the bistable magnetization of these bar nanomagnets behaves like macro-Ising spins[2-15]. Specially arranged interacting single-domain bar magnets in artificial spin ices enable the investigation of geometric frustration[2,3], emergent magnetic monopoles[6-8] and phase transitions[9-14] in a material-by-design approach. By coupling with the other functional materials, such as superconductors, these reconfigurable arrays of bar nanomagnets were used as reprogrammable magnetic potential landscapes to tailor the electronic properties of a hybrid device[15]. They were also used to tailor spin wave transformation for reconfigurable magnonic applications[16-20.] Recently, thermally active bar nanomagnets in artificial spin ice were adopted to demonstrate both deterministic and probabilistic computation[21]. More recently, utilizing single-domain bar nanomagnets with different magnetization reversal switching fields, a reprogrammable shape-morphing micromachine was designed, and showed great advances in controllability and functionality[22]. In all of the above, the magnetization reversal switching field of the bistable remnant magnetization is a crucial parameter for tuning the physical properties and functionalities of the assembled artificial systems. Therefore, the ability to modulate the magnetization switching field of single-domain bar nanomagnets would enable new routes for enhanced control and creation of new functionalities in artificial hybrid systems.

The magnetization reversal process of a nanomagnet is strongly affected by its magnetic anisotropy, which not only depends on the Fermi surface structure of the material, but also relies on the geometric shape of the magnet[23]. Adjusting the aspect ratio of a magnet, such as tuning its length, width and/or thickness, is a widely used method to control the magnetization reversal switching field, as recently demonstrated in magnetic shape-morphing microbots[22]. Here, using micromagnetic simulations we show the existence of an upper limit in the length of a bar nanomagnet, beyond which the magnetization reversal switching field cannot be tuned. This greatly limits the tuning of magnetization reversal by engineering aspect ratios, especially for a long nanomagnet. We circumvent this limitation by introducing a simple method to tailor the magnetization reversal process of a single-domain bar nanomagnet by shaping its end geometry. Our work provides an easily accessible method to tune the properties of mesoscopic magnetic systems based on single-domain bar nanomagnets.

**II. SIMULATION METHOD**

Micromagnetic simulations of the magnetization reversal processes on a bar nanomagnetic were carried out using MuMax3[24]. The material parameters of permalloy ($Ni_{80}Fe_{20}$) were used in our investigation because it is a soft magnetic alloy with very large permeability, and more importantly, it is one of the most widely used materials in related research[1-12,15-20]. The material parameters were set as standard values widely used for permalloy[25-30]: the saturation magnetization ($M_S$) is $8.6\times10^5$ A/m and the damping constant is 0.5 in order to quickly minimize the energy. The exchange stiffness ($A$) and the crystalline anisotropy constant ($K$) were $13\times10^{-12}$ J/m and 0 J/m³. The simulation cell size was 2.5nm × 2.5nm × 2.5nm.

## III. RESULTS AND DISCUSSION

We first investigated the magnetization reversal process by varying the aspect ratios of a standard, stadium-shaped nanomagnet with a semicircular geometry at both ends (Fig. 1a). The remanent magnetization spontaneously forms along the longitudinal direction of the bar under zero external magnetic field. The magnetic switching field $H_s$, i. e. the minimal magnetic field required to reverse the remanent magnetization, is determined from the sudden change of the magnetization hysteresis loop (Fig. 1d). The hysteresis loop is calculated by sweeping magnetic field along the long axis of the bar nanomagnet. The switching field $H_s$ decreases when the width is increased from 25 nm to 100 nm (Fig. 2b) and increases when the thickness is changed from 5 nm and 30 nm (Fig. 2c). Further tuning the switching field by width and/or thickness is not experimentally favorable, as fabricating nano-bars with width less than 25 nm requires very expensive ultrahigh resolution electron-beam lithography tools and a wider nanomagnet would transform from a single-domain state to a vortex state[1]. Furthermore, very thin nanomagnets have very low total magnetic moment which is not favored in the application of magnetic microbots[22]. In addition, very thin nanomagnets produce weak magnetic stray fields, thereby reducing the interactions between the nanomagnets and other proximal functional materials in hybrid devices. Fabricating nanomagnets using a thicker layer is challenging for the lift-off process. Moreover, a thicker sample changes the in-plane anisotropy of permalloy into an out-of-plane anisotropy, creating a complex domain and/or vortex structure in the bar nanomagnet (see Fig.S1). Due to all of the above reasons, tuning magnetization reversal by tailoring the nanomagnets' width and thickness has a variety of limitations.

Modulating the length of a bar nanomagnet with a moderate width and thickness seems to be an experimentally favorable way to control its magnetization reversal. To investigate the magnetization reversal's dependences on the nanomagnet's length, we fixed the nanomagnet's thickness and width to experimentally favorable values of 25 nm and 80 nm, respectively[3, 12] and varied its length $L$ between 160 nm and 1000 nm. A shorter nanomagnet ($L \leq 160$ nm) does not maintain the single-domain state and a vortex state emerges during the magnetization reversal process, as demonstrated in Video 1. We show the length dependence of the switching fields in Fig. 1e. Figure 1d displays the hysteresis loops of the bar nanomagnets with several selected lengths. One can see that the magnetization switching field increases when the length is increased from 160 nm to 480 nm. However, the switching field $H_s$ stays unchanged at lengths beyond 480 nm (Fig. 1e). This significantly limits the modulation of magnetization reversal of a single-domain bar nanomagnet by aspect ratios. Therefore, additional method is highly desired to further tune the magnetization reversal process, especially for a very long magnet.

The saturation of the switching field $H_s$ for a nanomagnet with a length beyond 480 nm suggests that the magnetization reversal process is not only controlled by the aspect ratios. Previous theoretical[31] and micromagnetic simulation[32-33] studies showed the anisotropy of the end of an elongated magnetic nanostructure plays crucial role. We plot the microscopic magnetic structures of the nanomagnets immediately before their reversal in the insets of Fig. 1d. Video 2 shows the reversal processes of several magnets with various length. The nucleation of the magnetization reversal begins at the two ends of the nanomagnet, and the magnetization at the central part of the nano-bar reverses instantly, following the magnetization reversal at the two ends. This indicates the magnetization reversal of the entire nano-bar is intimately dominated by its two ends. Previous investigations also showed that the magnetization properties of a submicrometer magnetic structure strongly depended on its shape[23,34-36]. Therefore, further tuning of the magnetization switching field of a bar nanomagnet could be realized by engineering the geometric shape of its two ends.

To demonstrate this approach, we investigated a simple nano-bar with semi-elliptical shaped ends (Fig. 2a). We fixed the vertical axis of the semi-ellipse to the width of the nano-bar and varied the length of the horizontal axis, $d$, as shown in Fig. 2a. The width and thickness of the nano-bar is 80 nm and 25 nm, respectively, while the total length was kept as 800 nm which is in the saturated magnetization reversal range $L > 480$ nm (Fig. 1e). This allows us to exam the effects from the end geometries on the entire bar by adjusting the horizontal axis, $d$, of the ellipse. As shown in Fig. 2b, hysteresis loops with different switching fields are obtained for bars with different axis lengths for the semi-ellipse. In Fig. 2c, we plot the switching field $H_s$ as a function of the axis length $d$. Increasing $d$ from 0 to 80 nm, reduces the switching field $H_s$ to a minimum value of about 83.5 mT. With further increasing $d$, $H_s$ increases rapidly up to a maximum value of 222.5 mT at $d$ = 320 nm. Video 3 demonstrates the evolution of the magnetization reversal dynamics with different $d$ values. We also simulate the $d$ dependence of the switching field for nano-bars with various lengths ranging from 300 nm to 1000 nm (supplemental Fig. S2). All the results indicate the nano-bar with semi-circle ends has the smallest switching field, and increasing the semi-ellipse axis length $d$ enhances the switching field more effectively than changing only the aspect ratio in Fig. 1.

The above results clearly indicate that the end geometry of a single-domain bar nanomagnet plays a critical role in the magnetization reversal process. While all the above investigations were performed on symmetric end geometries, it would be interesting to examine the case for asymmetrical end geometries. Subsequently, we carried out simulations for nanomagnets with elliptical axis lengths of $d_1$ and $d_2$ respectively, for its two ends, as illustrated in the inset of Fig. 3a. The overall size of the magnet bar was kept as 800 nm × 80 nm × 25 nm. In Fig. 3a, we plot the switching field $H_s$ as a function of both axis lengths $d_1$ and $d_2$. Figure 3b displays $d_1$ dependent switching field for several selected $d_2$ values. The results clearly show that the switching field can be effectively tuned by modifying the geometry of one end. Comparing the result of Fig. 3b with that from a symmetric nanomagnet displayed in Fig. 2c, we find the switching field of a nanomagnet with two different ends is determined by the end with the lower switching field. The inset of Fig. 3b displays the microscopic magnetic structure right before the switching for the nano-bar with asymmetric ends. The corresponding magnetization reversal process can be found in Video 4. We can clearly see that the nucleation of the magnetization reversal initiates at the end with the lower switching field. These simulation results indicate that effective and significant modulation of the magnetization reversal field can be realized by tuning the geometry of just one end of a single-domain bar nanomagnet.

In a real system consisting of bar nanomagnets, the applied external magnetic field may not always be directed along the long axes of all the bars. In all the aforementioned mesoscopic magnetic systems, such as in magnetic majority logic gates[1], artificial spin ices[2-21] and/or magnetic shape-morphing micromachines, the nanomagnets are patterned in various different orientations. Therefore, it is necessary to investigate the end geometry dependence of the magnetization reversal process of a single-domain nanomagnet under magnetic fields applied in different orientations. Figure 4a demonstrates the switching field as a function of the field angle $\theta_H$ and the axis lengths of the semi-ellipses at the ends. The size of the magnet bar was also kept as 800 nm × 80 nm × 25 nm. The result shows that the end geometry introduces a notable effect on the field-angle dependency of the magnetization reversal. We plot the field angle dependence of the switching fields for several selected $d$ values in Fig. 4b. For the nano-bar with large values of $d > 80$ nm, e.g., at $d = 200$ nm, the switching field decreases rapidly with increasing $\theta_H$ first, reaching a minimum at $\theta_H \approx 45°$ and then becoming larger at high field angle $\theta_H$ (green curve in Fig. 4b). That is, the field angle for easiest switching (minimal switching field) is around $\theta_H = 45°$. This result is consistent with the Stoner-Wohlfarth model for a single-domain ferromagnet[37-38]. The field angle for the minimal switching field decreases with the axis length $d$ and approaches ~20° when the end geometry becomes a semi-circle ($d = $ 80 nm). When further reducing the axis length of the semi-ellipses at the two ends, the angle dependent switching curves become more complex with a peak emerging at small angles ($\theta_H <$ 45°), as shown in Fig. 4b, which might be due to the competition between the shape anisotropy of the end segments and that of the center segment of the nano-bar. The difference in the switching fields for different $d$ is only shown at small field angles (roughly < 45º). The switching field curves under large field angles $\theta_H$ > 45º are nearly coincident. This indicates that the magnetization reversal process of the center segment of the nano-bar dominates over the end-segments when the magnetic field is tilted away from the long-axis of the bar.

## IV. CONCLUSTION

We have demonstrated a simple and effective way to engineer the magnetization reversal process of a single-domain bar nanomagnet without altering its overall size. By adjusting only the end geometries of a bar nanomagnet, the magnetization reversal switching field can be tuned over a wide field range. This method is more effective than tuning magnetization reversal by aspect ratios, especially for very long nano-bars in which the switching field is insensitive to the bar length beyond a certain span. We note that the 'effective' end-geometry of the nano-bars is not limited to semi-ellipse shapes, and that other geometries, such as triangular shapes (as demonstrated in Fig. S3), can also modulate the magnetization reversal. The ability to control the magnetization reversal of a single-domain bar magnet could lead to complex magnetization reversal behavior in coupled nanomagnet systems[29]. Fabricating arrays of nanomagnets with tailored magnetization reversal switching fields would enable modulation of phase transition temperatures and/or magnetic fields in artificial spin ices, which could greatly affect the access to ground state configurations as well as realizing novel excited low energy states. Creating artificial spin ices using nano-magnets with asymmetric end geometries may lead to novel phenomena by introducing local reversal barrier with designed defects. Nanomagnets with controllable magnetization reversals could also be used to modify phase coexistence[10] and domain wall frustration[39] in artificial magnetic systems. Our method would significantly simplify the design of magnetic shape-morphing micromachines, in which nanomagnets with different switching fields are required[22]. Engineered end-geometries of nanomagnets could also be used to control the operation process in magnetic majority logic gates[1], as well as to manipulate the tuning parameters of the current-driven magnetic logic devices[40].

**SUPPLEMENTARY MATERIAL**

Figures S1–S3 are present in the supplementary material. Videos 1–4 are also available as supplementary material.

**ACKNOWLEDGMENTS**

This work is supported by the National Key R&D Program of China (2018YFA0209002), the National Natural Science Foundation of China (61771235, 61971464, 61727805 and 11961141002), Jiangsu Excellent Young Scholar program (BK20200008) and Jiangsu Shuangchuang program. Z.-L.X. and W.K.K. acknowledge support from the U.S. Department of Energy, Office of Science, Basic Energy Sciences, Materials Sciences and Engineering. Z.-L.X. also acknowledges support from the National Science Foundation under Grant No. DMR-1901843. Y.D. acknowledges support from the Major Scientific Research Project of Zhejiang Lab (2019MB0AD01).

**DATA AVAILABILITY**

The data that support the findings of this study are available from the corresponding author upon reasonable request.

**Figures and Figure captions**

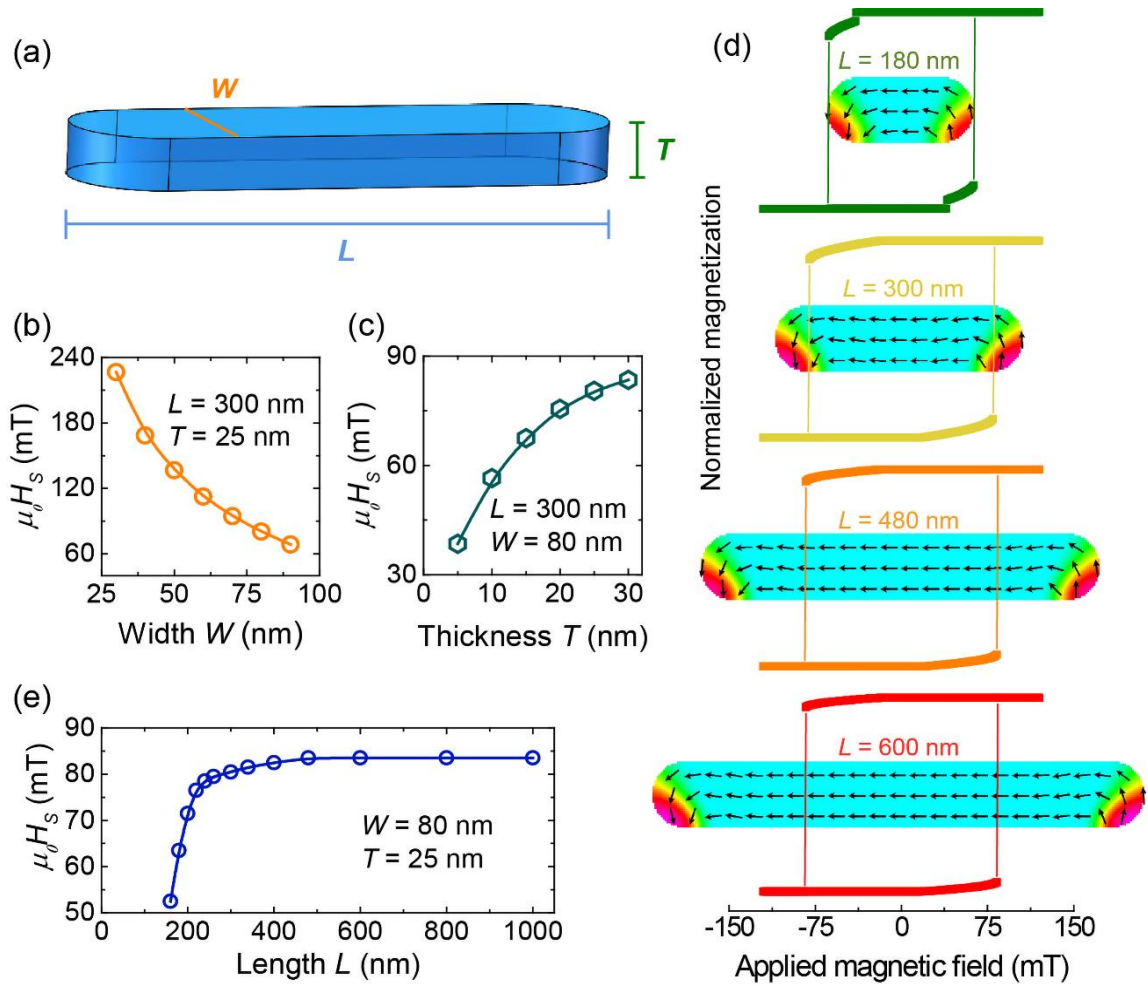

**Figure 1 | Dimension dependence of magnetization reversal of a bar nanomagnet. (a)** Schematic of a nanoscale bar magnet. **(b, c)** The reversal switching field $H_s$ of the bar magnet as a function of width **(b)** and thickness **(c)**, respectively. **(d)** Hysteresis loops of the bar magnets of various lengths. **(e)** Length dependence of switching field for a nanomagnet with fixed width, W, and thickness, T, as shown in figure **(a)**.

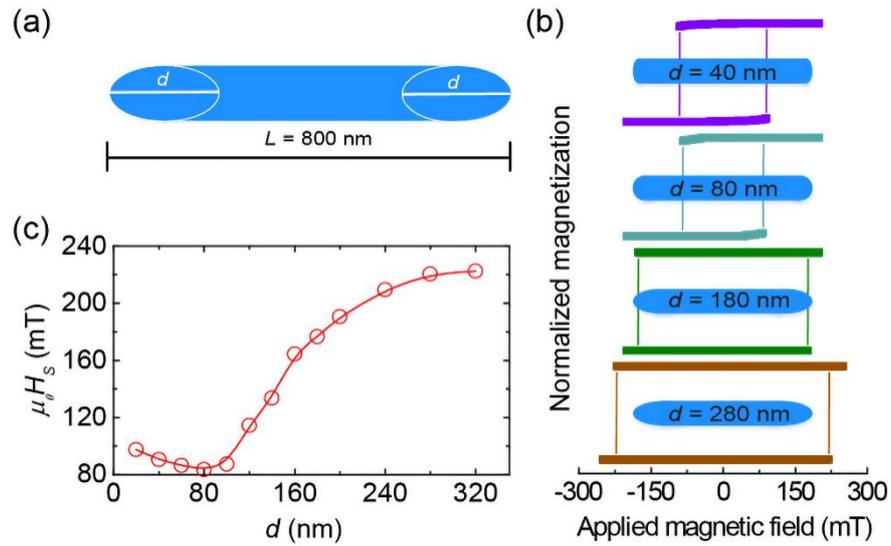

**Figure 2 | Tuning magnetization reversal with symmetric end-geometries. (a)** Schematic of a bar nanomagnet with semi-elliptical ends. The length and width of the bar are 800 nm and 80 nm, respectively. The elliptical axis length d is the parameter for tuning the end-geometry. **(b)** Hysteresis loops of the bar magnets with various identical elliptical axis lengths at the two ends. **(c)** End shape (or axis length of the semi-ellipse) dependence of the switching fields.

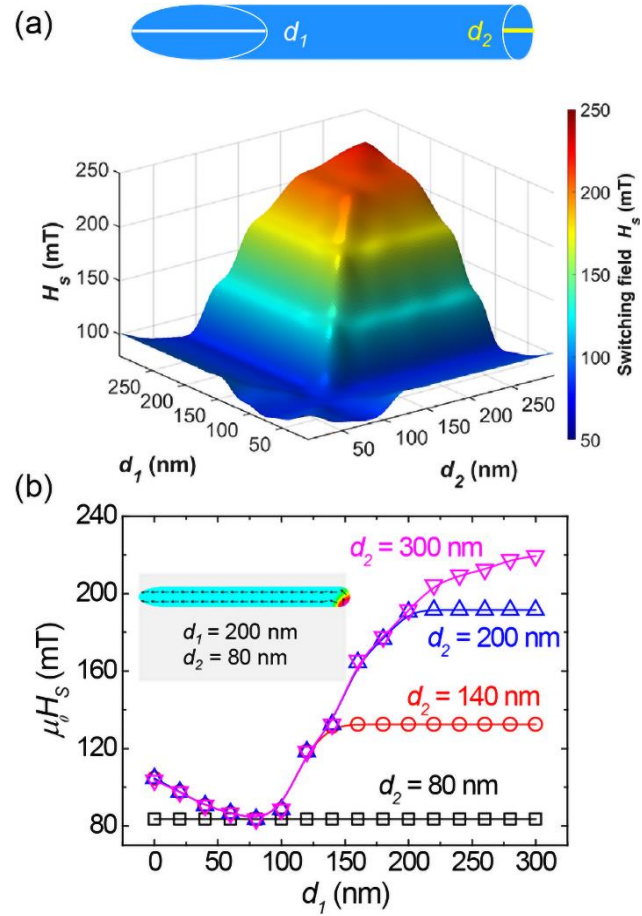

**Figure 3 | Regulating magnetization reversal with asymmetric end-geometries. (a)** Switching field, $H_s$, as a function of the semi-ellipse axis lengths, $d_1$ and $d_2$, for the two end-geometries of the bar as defined in the top inset. The length and width of the bar are respectively fixed at 800 nm and 80 nm. **(b)** Switching field as a function of axis length $d_1$, depicting curves with several values of axis length $d_2$. The inset is the micromagnetic structure immediately before the magnetization reversal switching of a bar nanomagnet with different axis lengths at its two ends.

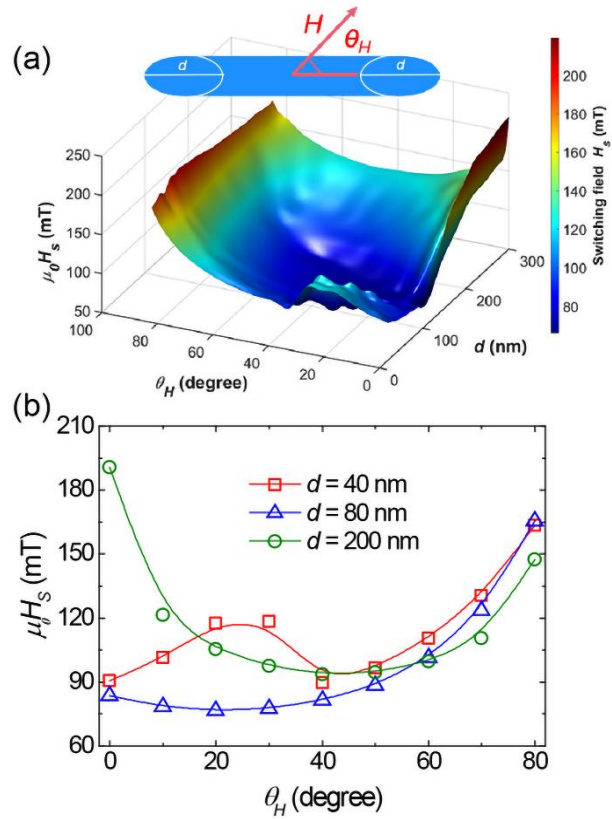

**Figure 4 | Effects of magnetic field orientation on the magnetization reversal of a bar magnet. (a)** Switching field as a function of field orientation $\theta$ and axis length $d$ of the semi-ellipses at the symmetric bar ends. **(b)** Field orientation dependence of switching field at several selected axis lengths $d$ of the semi-ellipses at the bar ends.